\documentstyle[12pt]{article}

\makeatletter
\renewcommand\thesubsection{\thesection.\@arabic\c@subsection}
\makeatother

\newcommand{\sect}[1]{\setcounter{equation}{0}\section{#1}}

\newcommand {\beq}{\begin{equation}}
\newcommand {\eeq}{\end{equation}}
\newcommand {\beqa}{\begin{eqnarray}}
\newcommand {\eeqa}{\end{eqnarray}}         
\newcommand {\beqs}{\begin{eqnarray*}}
\newcommand {\eeqs}{\end{eqnarray*}}
\newcommand {\bds}{\begin{displaymath}}
\newcommand {\eds}{\end{displaymath}}
\newcommand {\n}{\nonumber\\}

\newcommand {\bebb}{\begin{thebibliography}}
\newcommand {\eebb}{\end{thebibliography}}      
\newcommand {\bbit}{\bibitem}

\def\pd{\prod}

\def\ra{\rangle}

\def\dg{\dagger}

\def\dz{\frac{d}{dz}}

\begin{document}




\vskip 1cm

\begin{center}
{\Large\bf On the 2-mode and $k$-photon quantum Rabi models}

\vspace{0.5cm}

{\large Yao-Zhong Zhang}
\vskip.1in

{\em School of Mathematics and Physics, The University of Queensland, \\
Brisbane, Qld 4072, Australia}

{\em CAS Key Laboratory of Theoretical Physics, Institute of Theoretical Physics,\\
Chinese Academy of Sciences, Beijing 100190, China}

\end{center}

\date{}



\begin{abstract}
By mapping the Hamiltonians of the two-mode and 2-photon Rabi models to differential operators in suitable Hilbert spaces of entire functions, we prove that the two models possess entire and normalizable wavefunctions 
in the Bargmann-Hilbert spaces only if the frequency $\omega$ and coupling strength $g$ satisfy certain constraints.
This is in sharp contrast to the quantum Rabi model for which entire wavefunctions always exist. 
For model parameters fulfilling the aforesaid constraints we determine transcendental equations whose roots give the regular energy eigenvalues of the models. 
Furthermore, we show that for $k\geq 3$  the $k$-photon Rabi model does not possess wavefunctions 
which are elements of the Bargmann-Hilbert space for all non-trivial model parameters. 
This implies that the $k\geq 3$ case is not diagonalizable, unlike its RWA cousin, the $k$-photon Jaynes-Cummings
model which can be completely diagonalized for all $k$.   

\end{abstract}

\vskip.1in

{\tt PACS numbers}:  03.65.Ge, 02.30.Ik, 42.50.Pq.








\sect{Introduction}

Since Braak's breakthrough work on the quantum Rabi model \cite{Braak11}, there have been renewed interest in analytic solutions of this model \cite{Moroz12,Chen12,Zhong13,Moroz13,Moroz14,Tomka14,Batchelor15a,Batchelor15b} 
and its multi-quantum and multi-level generalizations \cite{Albert12,Zhang13a,Zhang13c,Braak13,Zhang14,Chen14,Peng15}. 
Prior to the works on analytic solutions, the $k$-photon Rabi model was investigated mainly via numerical diagonalization techniques (see e.g. \cite{Lo98,Ng99}).  
It was argued in \cite{Lo98} that this model is non-diagonalizable and thus ill-defined for $k>2$. This is believed to be 
related to the impossibility of higher order squeezing in the traditional sense in that $i\xi^* a^k-i\xi(a^\dagger)^k$, 
with $\xi$ being complex parameter and $a^\dagger~(a)$ boson creation (annihilation) operators, 
is not essentially self-adjoint for $k>2$ \cite{Gorska14}. 
For $k=2$ (the 2-photon Rabi case), the constraint condition $|2g/\omega|<1$, with $g$ being the coupling strength and $\omega$ the frequency, became present
\cite{Zhang13a,Zhang13c} if the (Bogoliubov) transformations used in solving the model are to make sense. The situation is the same in the 2-mode Rabi case, but with the model parameters satisfying the constraint $|g/\omega| <1$ \cite{Zhang13a,Zhang13c}. 
These results indicate that the 2-mode and $k$-photon Rabi models are qualitatively quite different from the quantum Rabi model.  

The aim of this paper is to provide an independent proof to the following (somewhat unexpected) results based on the application of the Bargmann-Hilbert spaces: (i) the 2-mode and 2-photon Rabi models are defined in suitable Hilbert spaces of entire analytic functions only if $\omega, g$ satisfy the constraints $|g/\omega| <1$ and $|2g/\omega|<1$, respectively; 
and (ii) for $k\geq 3$ the $k$-photon Rabi model is not defined for all non-trivial model parameters to the extent that it does not possess normalizable entire wavefunctions in the Bargmann-Hilbert space. 
We build our proof from the algebraization and partial diagonalization of the boson degrees of freedom. 
For the 2-mode and 2-photon Rabi models with model parameters satisfying the aforesaid constraints, 
we also determine transcendental equations whose roots  give the regular energy eigenvalues of the models.


\sect{Two-mode Rabi model}\label{two-mode}
The Hamiltonian of the two-mode quantum Rabi model reads
\beq
H=\omega(a_1^\dagger a_1+a_2^\dagger a_2)+\Delta\sigma_z+g\,\sigma_x(a_1^\dagger a_2^\dagger+a_1 a_2),
  \label{2-mode-RabiH1}
\eeq
where $a^\dagger~(a)$ are creation (annihilation) operators of boson modes with frequency $\omega$
(we have assumed that the boson modes are degenerate with the same frequency),
$\sigma_z, \sigma_x$ are Pauli matrices describing two atomic levels
separated by energy difference $2\Delta$, and $g$ is the spin-boson interaction strength. 
The Hilbert space of the model, ${\cal H}_b\otimes {\cal C}^2$, is infinite dimensional, 
where ${\cal H}_b$ is the Hilbert space of the boson and ${\cal C}^2$ is the spin space.

\subsection{Partial diagonalization}
The bosonic fields allow an algebraization. Introduce the operators $K_\pm, K_0$,
\beq
K_+=a_1^\dagger a_2^\dagger,~~~~K_-=a_1 a_2,~~~~K_0=\frac{1}{2}(a_1^\dagger a_1+a_2^\dagger a_2+1).
\label{2-mode-Rabi-boson}
\eeq
These operators form the usual $su(1,1)$ Lie algebra. 
This algebra has an infinite-dimensional unitary irreducible representation 
(i.e. the positive discrete series) ${\cal D}^\kappa$ labeled by the Bargmann index $\kappa$. 
For the two-mode bosonic realization (\ref{2-mode-Rabi-boson}) of $su(1,1)$ that we require here
the Bargmann index $\kappa$ can take any positive integers or half-integers, 
i.e. $\kappa=1/2, 1, 3/2,\cdots$. Thus by means of the $su(1,1)$ representation we have decomposed 
the Fock-Hilbert space ${\cal H}_b$ of the system into the direct sum of infinite number of 
subspaces ${\cal H}_b^\kappa$ labelled by $\kappa=1/2, 1, 3/2, \cdots$. Within the subspace 
${\cal H}_b^\kappa$ 
the Hamiltonian is given by
\beq
H^{(\kappa)}=2\omega\left(K_0-\frac{1}{2}\right)+\Delta\sigma_z+g\sigma_x(K_++K_-).\label{2-mode-RabiH2}
\eeq
In other words, the algebraization allows a partial diagonalization of the Hamiltonian (\ref{2-mode-RabiH1})
by bringing it into block-diagonal form,
\beq
H=\bigoplus_\kappa\,H^{(\kappa)},\label{2-modeH1-decomposition}
\eeq
where $H^{(\kappa)}$  acts in the mutually orthogonal subspaces ${\cal H}_b^\kappa\otimes {\cal C}^2$ 
with fixed $\kappa$.
Thus the problem of diagonalizing the two-mode Rabi model (\ref{2-mode-RabiH1}) is reduced to
that of diagonalizing each $H^{(\kappa)}$ in the corresponding subspace 
${\cal H}_b^\kappa\otimes {\cal C}^2$  separately.

The Hamiltonian $H^{(\kappa)}$ possess a ${\cal Z}_2$ symmetry (parity), $P\,H^{(\kappa)}\,P=H^{(\kappa)}$, 
where $P=e^{i\pi(K_0-\kappa)}\otimes\sigma_z$ is the parity operator in the subspace 
${\cal H}_b^\kappa\otimes {\cal C}^2$. Thus each ${\cal H}_b^\kappa\otimes {\cal C}^2$ splits into two
invariant subspaces ${\cal H}_b^\kappa\otimes |\pm\ra$ labelled by the eigenvalues $\pm 1$ of $P$. 
This parity invariance can be used to partially diagonalize $H^{(\kappa)}$. 
This is seen as follows. Define the unitary operator
\beq
U=\frac{1}{\sqrt{2}}\left(
\begin{array}{cc}
1 & 1 \\
T & -T\\
\end{array}
\right), ~~~~~~
T=e^{i\pi(K_0-\kappa)}. 
\eeq
Working in a representation defined by $\sigma_x$ diagonal, we have 
\beq
U^\dagger\,H^{(\kappa)}\,U=\left(
\begin{array}{cc}
H^{(\kappa)}_+ & 0\\
0 & H^{(\kappa)}_-\\
\end{array}
\right),
\eeq
where for fixed $\kappa$
\beq
H^{(\kappa)}_\pm=2\omega\left(K_0-\frac{1}{2}\right)+g(K_++K_-)\pm \Delta\,T.\label{2-mode-RabiH3}
\eeq
act in two mutually orthogonal subspaces ${\cal H}^\kappa_b\otimes |\pm\ra$ with fixed parity.

\subsection{Wavefunctions and constraints for $\omega$ and $g$}
In the same way as the differential realization of boson operators in a Hilbert space of entire analytic functions
of growth $(\frac{1}{2}, 1)$ \cite{Bargmann61}, we can represent the continuous boson degree of freedom 
($K_{\pm, 0}$) as differential operators in a Hilbert space of entire functions 
of growth $(1,1)$ \cite{Barut71}. 

Let ${\cal B}_\kappa$ be the Hilbert space associated with $\kappa$ whose elements are entire analytic functions,  
called Bargmann-Hilbert space ${\cal B}_\kappa$. Its inner product is defined by \cite{Barut71}
\beq
(f,g)_\kappa=\int\,\overline{f(z)}\,g(z)\,d\mu_\kappa(z),~~~~~~d\mu_\kappa(z)=\frac{4}{\pi}|z|^{2\kappa-1}
K_{\frac{1}{2}-2\kappa}(2|z|)\,dxdy,\label{norm-for-2-mode}
\eeq
where $d\mu_\kappa(z)$ is the measure and $K_\nu(z)$ is the modified Bessel function of the third kind which has the Mellin transform, 
$\int_0^\infty 2\xi^{\alpha+\beta}K_{\alpha-\beta}(2\xi^{1/2})\xi^{s-1}d\xi=\Gamma(s+2\alpha)\Gamma(s+2\beta)$.  
Then $f(z)$ belongs to ${\cal B}_\kappa$ if and only if $||f||_\kappa^2=(f,f)_\kappa<\infty$. 
It is now not difficult to see that if $f(z)=\sum_{n=0}^\infty\,c_n z^n$, then
\beq
||f||_\kappa^2=\sum_{n=0}^\infty \,|c_n|^2\,n!\left(n+2\kappa-1\right)!\label{2-mode-entireness-test}.
\eeq
Every set of coefficients $c_n$ for which the sum on the right hand side converges defines an entire function
$f(z)\in {\cal B}_\kappa$. An orthonormal set of basis vectors in  ${\cal B}_\kappa$ is given
by the monomials $\left\{{z^n}/{\sqrt{n!(n+2\kappa-1)!}}\right\}$. In this basis
$K_\pm, K_0$ (\ref{2-mode-Rabi-boson}) are realized as the single-variable differential operators \cite{Zhang13a,Barut71}
\beq
K_0=z\frac{d}{dz}+\kappa,~~~~K_+=z,~~~~K_-=z\frac{d^2}{dz^2}+2\kappa\frac{d}{dz}.
      \label{su11-diff-rep-2mode}
\eeq
The operator $T$ can be realized as $T=e^{i\pi\,z\dz}$, which acts on elements $f(z)$ of ${\cal B}_\kappa$ as 
$(T\,f)(z)=f(-z)$. 

By means of the differential representation (\ref{su11-diff-rep-2mode}), we can express the Hamiltonian 
(\ref{2-mode-RabiH3}) as the differential operator in ${\cal B}_\kappa$
\beq
H^{(\kappa)}_\pm=2\omega\left(z\frac{d}{dz}+\kappa-\frac{1}{2}\right)\pm \Delta\, e^{i\pi \,z\frac{d}{dz}}+g\left(z+z\frac{d^2}{dz^2}
   +2\kappa\frac{d}{dz}\right).   \label{2-mode-Rabi-diffH}
\eeq
The corresponding time-independent Schr\"odinger equations are
\beq
\left\{gz\frac{d^2}{dz^2}+2(\omega z+g\kappa )\frac{d}{dz}\pm \Delta\, e^{i\pi \,z\frac{d}{dz}}
+gz+2\omega\left(\kappa-\frac{1}{2}\right)-E_\pm\right\}\phi^\pm(z)=0.\label{2-mode-Schroedinger}
\eeq
Here we have written $E_\pm$ since in general the spectra of $H^{(\kappa)}_\pm$ are not the same.
As in the Rabi case \cite{Schweber67}, $E_\pm$ belong to the spectra of $H^{(\kappa)}_\pm$ if and only if for 
these values of $E_\pm$ solutions to the above differential equations are entire functions belonging to 
${\cal B}_\kappa$. In other words, we are seeking solutions of the form
\beq
\phi^\pm(z)=\sum_{n=0}^\infty S_n^\pm(E_\pm)\, z^n, \label{2-mode-series-solution}
\eeq
which converge in the entire complex plane and are elements of ${\cal B}_\kappa$. 

Substituting (\ref{2-mode-series-solution}) into (\ref{2-mode-Schroedinger}), 
we obtain the 3-step recurrence relation,
\beqa
&&S_1^\pm+C_0^\pm\,S_0^\pm=0,\n
&&S_{n+1}^\pm+C_n^\pm\,S_n^\pm+D_n^\pm\,S_{n-1}^\pm=0,~~~~n\geq 1,\label{2-mode-3-step}
\eeqa
where
\beq
C_n^\pm=\frac{\pm(-1)^n\,\Delta-E_\pm+2\omega\left(n+\kappa-\frac{1}{2}\right)}
  {g(n+1)(n+2\kappa)},~~~~~~
D_n^\pm=\frac{1}{(n+1)(n+2\kappa)}.\label{coefficients CD}
\eeq
The coefficients $C_n^\pm, D_n^\pm$ have the behavior
\beq
C_n^\pm\sim \frac{2\omega}{g}\,n^{-1},~~~~ D_n^\pm\sim n^{-2}\label{2-mode-asymptotoc}
\eeq
when $n\rightarrow\infty$. 
Thus the asymptotic structure of solutions to the $n\geq 1$ part of (\ref{2-mode-3-step}) 
depends on the Newton-Puiseux diagram formed with the points $P_0(0,0), P_1(1,-1), P_2(2,-2)$ \cite{Gautschi67}. 
The characteristic equation of the 3-term recurrence relation is given by
$t^2+\frac{2\omega}{g}t+1=0$, which has two solutions
$t_{1,2}=-\frac{\omega}{g}\pm\sqrt{\frac{\omega^2}{g^2}-1}$.
Thus we have two cases to consider. 
\vskip.2in
\noindent\underline{\large\bf Case (1)}: $\left|\frac{g}{\omega}\right|<1$. 
In this case, we have two distinct
real roots $t_1=\frac{\omega}{g}\left[-1+\sqrt{1-(g/\omega)^2}\right]$, 
$t_2=-\frac{\omega}{g}\left[1+\sqrt{1-(g/\omega)^2}\right]$, and $|t_1|<|t_2|$. 
The Perron-Kreuser theorem (i.e. Theorem 2.3 of \cite{Gautschi67}) 
gives the asymptotic behaviour of two linearly independent
solutions $S^\pm_{n,r}$, 
\beq
\lim_{n\rightarrow\infty}\frac{S^\pm_{n+1,r}}{S^\pm_{n,r}}\sim t_r\,n^{-1},~~~~r=1,2.
\eeq
So solution $S^\pm_{n,1}$ is minimal while $S^\pm_{n,2}$ is dominant. 
{}From (\ref{2-mode-entireness-test}), for an entire solution in ${\cal B}_\kappa$ the sum
\beq
\sum_{n=0}^\infty\,|S^\pm_n|^2\,n!\,(n+2\kappa-1)!\label{entireness-test-for-2mode-Rabi}
\eeq
must converge. Using the ratio test,
\beq
\lim_{n\rightarrow \infty}\frac{|S^\pm_{n+1}|^2\,(n+1)!\,(n+2\kappa)!}{|S^\pm_n|^2\,n!\,(n+2\kappa-1)!}
  =|t_r|^2
\eeq
It is easily seen that $|t_2|^2>1$. We can show that $|t_1|^2<1$. Indeed, if we had 
$|t_1|^2\geq 1$, then we would end up with $\sqrt{1-\left|{g}/{\omega}\right|}
\geq \sqrt{1+\left|{g}/{\omega}\right|}$ which is impossible for the non-trivial case $g\neq 0$. 

It follows that the sum (\ref{entireness-test-for-2mode-Rabi}) converges for the minimal solution 
$S^\pm_{n,1}$ and thus the corresponding wavefunctions (\ref{2-mode-series-solution}) 
are elements of ${\cal B}_\kappa$.

\vskip.1in
\noindent\underline{\large\bf Case (2)}: $\left|\frac{g}{\omega}\right|\geq 1$. 
In this case, the two roots $t_{1,2}$ are
complex conjugate to each other and $|t_1|=|t_2|=1$. Applying the Perron-Kreuser theorem, we have
\beq
\lim_{n\rightarrow\infty}\,{\rm sup}\,\left(|S^\pm_n|\,n!\right)^{\frac{1}{n}}=1
\eeq
for all non-trivial solutions of the 2nd equation of (\ref{2-mode-3-step}). Thus given $\epsilon>0$,
there exists $N(\epsilon)\in {\bf N}$ and an infinite set $I$ of indices $\ell>N(\epsilon)$
such that $\left(|S^\pm_\ell|\,\ell!\right)^{\frac{1}{\ell}}> 1-\epsilon$, i.e.
$|S^\pm_\ell|>(1-\epsilon)^\ell/\ell!$. So we have
\beqa
\sum_{n=0}^\infty\,|S^\pm_n|^2\,n!\,(n+2\kappa-1)! &\geq& \sum_{\ell\in I}\,|S^\pm_\ell|^2\,\ell!\,(\ell+2\kappa-1)!\n
&>& \sum_{\ell\in I}\,\left(1-\epsilon\right)^{2\ell}\,\frac{(\ell+2\kappa-1)!}{\ell!}.
\label{inequality1}
\eeqa
Noting that when $\ell\rightarrow\infty$, $\epsilon\rightarrow 0$, we have
\beq
\lim_{\ell\rightarrow \infty}\left(1-\epsilon\right)^{2\ell}\,\frac{(\ell+2\kappa-1)!}
   {\ell!}\neq 0.
\eeq
This means the series on the right hand side of (\ref{inequality1}) diverges.
By the comparison test, the series on the left hand side of (\ref{inequality1}), i.e.
the sum (\ref{entireness-test-for-2mode-Rabi}) diverges  for all non-trivial solutions of the
3-term recurrence relations. Thus  when $\left|\frac{g}{\omega}\right|\geq 1$ the two-mode Rabi model has no
entire wavefunctions which belong to ${\cal B}_\kappa$.

\subsection{Energy spectrum}
We now proceed to find regular energy eigenvalues $E_\pm$ corresponding to $|g/\omega|<1$ and the
minimal solutions $S_n^{\pm min}\equiv S^\pm_{n,1}$ (and thus to the entire wavefunctions $\phi_\pm(z)$).
We follow a procedure presented in \cite{Leaver86,Moroz12,Zhang13c,Zhang14} that uses the relationship between minimal
solutions and infinite continued fractions \cite{Gautschi67}. 

By the Pincherle theorem (i.e. Theorem 1.1 of \cite{Gautschi67}), 
the ratios of successive elements
of the minimal solutions $S^{\pm min}_n$ can be expressed as continued fractions,
\beq
{\cal S}^\pm_{n}=\frac{S^{\pm min}_{n+1}}{S^{\pm min}_n}=-\frac{D^\pm_{n+1}}{~C^\pm_{n+1}-}\,
  \frac{D^\pm_{n+2}}{~C^\pm_{n+2}-}\,\frac{D^\pm_{n+3}}{~C^\pm_{n+3}-}\,\cdots,   \label{2-mode-continued-fraction}
\eeq
which for $n=0$ reduces to
\beq
{\cal S}^\pm_{0}=\frac{S^{\pm min}_{1}}{S^{\pm min}_0}=-\frac{D^\pm_{1}}{~C^\pm_{1}-}\,\frac{D^\pm_{2}}{~C^\pm_{2}-}\,
\frac{D^\pm_{3}}{~C^\pm_{3}-}\,\cdots.   \label{2-mode-continued-fraction1}
\eeq
The ratios ${\cal S}^\pm_0=\frac{S^{\pm min}_1}{S^{\pm min}_0}$ involve $S^{\pm min}_n$, 
although the above continued fraction expressions are obtained from the 2nd equation 
of (\ref{2-mode-3-step}), i.e the recurrence (\ref{2-mode-3-step}) for $n\geq 1$. 
On the other hand, 
for single-ended sequences such as those appearing in the infinite series expansions (\ref{2-mode-series-solution}), 
the ratios ${\cal S}^\pm_0=\frac{S^{\pm min}_1}{S^{\pm min}_0}$ of the first two terms of 
a minimal solution are unambiguously fixed by  the $n=0$ part of 
the recurrence (\ref{2-mode-3-step}), that is,
\beq
{\cal S}^\pm_0=-C^\pm_0=\frac{1}{2g\kappa}\left[E_\pm\mp\,\Delta-2\omega\left(\kappa-\frac{1}{2}\right)\right]. 
   \label{2-mode-continued-fraction2}
\eeq
In general, (\ref{2-mode-continued-fraction1})  and (\ref{2-mode-continued-fraction2}) can not be both satisfied
 i.e. the ${\cal S}^\pm_0$ computed from (\ref{2-mode-continued-fraction1}) are not the same as that from 
(\ref{2-mode-continued-fraction2}) 
for arbitrary values of the recurrence coefficients $C_n$ and $D_n$.  
Thus general solutions to the recurrence (\ref{2-mode-3-step}) are dominant and are usually
generated by simple forward recursion from a given value of $S_0^\pm$. 
Physical meaningful solutions are those that are entire and normalizable with respect to the the Bargmann-Hilbert space
norm given in (\ref{norm-for-2-mode}). They can be obtained if $E_\pm$ can be adjusted so that equations 
(\ref{2-mode-continued-fraction1}) and (\ref{2-mode-continued-fraction2}) are both satisfied. 
Then the resulting solution sequences ${\cal S}_n^\pm(E_\pm)$ will be purely minimal and
the corresponding power series expansions (\ref{2-mode-series-solution}) will converge in the whole complex plane
and be elements of ${\cal B}_\kappa$.
Equating the right hand sides of (\ref{2-mode-continued-fraction1}) and (\ref{2-mode-continued-fraction2})
yields implicit continued fraction equations for the regular spectrum $E_\pm$,
\beq
0=C^\pm_0-\frac{D^\pm_{1}}{~C^\pm_{1}-}\,\frac{D^\pm_{2}}{~C^\pm_{2}-}\,
\frac{D^\pm_{3}}{~C^\pm_{3}-}\,\cdots.   \label{2-mode e-value eqn}
\eeq
Here $C^\pm_n, D^\pm_n$ are defined as functions of $E_\pm$ in 
(\ref{coefficients CD}), and thus (\ref{2-mode e-value eqn}) contain $E_\pm$ as parameters. These are transcendental equations for the determination of the regular energies $E_\pm$ of the two-mode Rabi model.
Only for the denumerable infinite values of $E_\pm$ which are the roots of (\ref{2-mode e-value eqn}), 
do the power series (\ref{2-mode-series-solution})
give convergent and normalizable solutions to the differential equations (\ref{2-mode-Schroedinger}).



\sect{$k$-photon Rabi model}\label{k-photon}
The $k$-photon Rabi model is a natural generalization of the quantum Rabi model. Its Hamiltonian 
is given by 
\beq
H=\omega a^\dagger a+\Delta\,\sigma_z+g\,\sigma_x\left[(a^\dagger)^k+a^k\right],\label{k-photonH1}
\eeq
where $k=1,2,\cdots$ is a positive integer, $a^\dagger~(a)$ are creation (annihilation) operators
of a boson mode with frequency $\omega$, $\sigma_z, \sigma_x$ are Pauli matrices describing two atomic levels
separated by energy difference $2\Delta$, and $g$ is the coupling strength.
As in the 2-mode case, we will let ${\cal H}_b\otimes {\cal C}^2$ denote the Hilbert space of the $k$-photon Rabi model, where ${\cal H}_b$ is the Hilbert space of boson number states and ${\cal C}^2$ is the spin space.
The $k=1$ case of  (\ref{k-photonH1}) gives the Hamiltonian of the Rabi model.

\subsection{Partial diagonalization} 
The boson fields allow an algebraization. Introduce three operators
\beq
Q_+=\frac{1}{\sqrt{k^k}}(a^{\dg})^k ,~~~~
Q_-=\frac{1}{\sqrt{k^k}}a^k , ~~~~
Q_0 =\frac{1}{k} \left(a^{\dg}a+ \frac{1}{k} \right).\label{su11-poly-boson}
\eeq
These operators form a polynomial algebra given in \cite{Lee10}.
The polynomial algebra has an infinite-dimensional unitary irreducible
representation labelled by the parameter $q$, which for $k=2$ reduces to the well-known positive 
discrete series of $su(1,1)$. For the single-mode bosonic realization 
(\ref{su11-poly-boson}), the parameter $q$  takes the $k$ allowed values \cite{Lee10},
\beq
q= \frac{1}{k^2} ,~ \frac{k+1}{k^2},~\frac{2k+1}{k^2} ,\cdots, ~\frac{(k-1)k+1}{k^2}.\label{q-values}
\eeq
Thus by the algebraization of the bosonic field, we can decompose the Hilbert space ${\cal H}_b$  
into a direct sum of $k$ independent subspaces ${\cal H}^q_b$ labelled by $q$.
Within the subspace ${\cal H}_b^q$ 
with fixed $q$, where $n=0,1,2,\cdots$, the Hamiltonian is given by
\beq
H^{(q)}=k\omega\left({Q}_0 -\frac{1}{k^2}\right)+\Delta\,\sigma_z
    + g\sqrt{k^k}\,\sigma_x\left( {Q}_+ + {Q}_- \right).\label{k-photonH2}
\eeq
In other words, the algebraization allows a partial diagonalization of the
Hamiltonian (\ref{k-photonH1}) by bringing it into block-diagonal form, 
\beq
H=\bigoplus_q\,H^{(q)},\label{k-photonH1-decomposition}
\eeq
where $H^{(q)}$  acts in $k$ mutually orthogonal subspaces ${\cal H}_b^q\otimes {\cal C}^2$ with fixed $q$.
Thus the problem of diagonalizing the $k$-photon Rabi model (\ref{k-photonH1}) is reduced to
that of diagonalizing each $H^{(q)}$ in the corresponding subspace 
${\cal H}_b^q\otimes {\cal C}^2$  separately.

The Hamiltonian $H^{(q)}$ possess a ${\cal Z}_2$ symmetry (parity), $P_q\,H^{(q)}\,P_q=H^{(q)}$, where
$P_q=e^{i\pi(Q_0-q)}\otimes\sigma_z$ is the parity operator in the subspace 
${\cal H}_b^q\otimes {\cal C}^2$. Thus each ${\cal H}_b^q\otimes {\cal C}^2$ splits into two
invariant subspaces ${\cal H}_b^q\otimes |\pm\ra$ labelled by the eigenvalues $\pm 1$ of $P_q$. 
This parity invariance can be used to partially diagonalize $H^{(q)}$.
This is seen as follows. Define the unitary operator
\beq
U=\frac{1}{\sqrt{2}}\left(
\begin{array}{cc}
1 & 1 \\
T & -T\\
\end{array}
\right), ~~~~~~
T=e^{i\pi(Q_0-q)}. 
\eeq
Working in a representation defined by $\sigma_x$ diagonal, we have 
\beq
U^\dagger\,H^{(q)}\,U=\left(
\begin{array}{cc}
H^{(q)}_+ & 0\\
0 & H^{(q)}_-\\
\end{array}
\right),
\eeq
where for fixed $q$
\beq
H^{(q)}_\pm=k\omega\left({Q}_0 -\frac{1}{k^2}\right)+g\sqrt{k^k}\left( {Q}_+ + {Q}_- \right)\pm \Delta\,T\label{k-photonH3}
\eeq
 act in two mutually orthogonal subspaces ${\cal H}^q_b\otimes |\pm\ra$ 
with fixed parity.

\subsection{Bargmann-Hilbert space and 3-term recurrence relations}

The continuous boson degree of freedom $Q_{\pm, 0}$ of the $k$-photon Rabi model can be realized 
as differential operators in a Hilbert space of entire functions of growth $(1,1)$ \cite{Zhang13b}. Let
${\cal B}_q$ denote the Hilbert space associated with index $q$, called Bargmann-Hilbert space ${\cal B}_q$.
Similar to the two-mode case, the inner product in ${\cal B}_q$ is defined by
\beq
(f,g)_q=\int\overline{f(z)}\,g(z)\,d\mu_q(z),~~~~~~d\mu_q(z)=\frac{1}{\pi}|z|^{2(q-\frac{k(k-1)+1}{k^2})}
e^{-|z|^{2/k}}\,dxdy.\label{norm-for-k-photon}
\eeq
$f$ belongs to ${\cal B}_q$ if and only if $||f||_q^2=(f,f)_q<\infty$.
Now let $f(z)$ be an entire function with the power series $f(z)=\sum_{n=0}^\infty c_n\,z^n$. Then it can be easily checked using $\Gamma(s)=\int^\infty_0 \xi^{s-1}e^{-\xi}d\xi$ for Re$(s)>0$ that in terms of the expansion coefficients
\beq
||f||_q^2=\sum_{n=0}^\infty\,|c_n|^2\,[k(n+q-1/k^2)]!\label{k-photon-entireness-test}
\eeq
Every set of coefficients $c_n$ for which the sum on the right hand side converges defines an entire analytic function 
$f\in {\cal B}_q$.
An orthonormal set of basis vectors in ${\cal B}_q$ is given by the monomials 
$\left\{z^n/\sqrt{[k(n+q-1/k^2)]!}\right\}$.
In this basis, $Q_{\pm, 0}$ are represented as single-variable differential operators \cite{Zhang13b},
\beqa
{Q}_0 &=& z\frac{d}{dz}+ q, ~~~~~~{Q}_+ =\frac{z}{\sqrt{k^k}}, \n
{Q}_- &=& z^{-1}\,\sqrt{k^k} \pd_{j=1}^{k}  \left(z\dz+q- \frac{(j-1)k+1}{k^2} \right).
\label{su11-poly-diff}
\eeqa
Moreover the operator $T$ can be realized as $T=e^{i\pi\,z\dz}$, which acts on elements $f(z)$ of ${\cal B}_q$  as 
$(T\,f)(z)=f(-z)$.

Using the differential realization  we can equivalently write (\ref{k-photonH3}) 
as the single-variable differential operator in ${\cal B}_q$,
\beqa
H^{(q)}_\pm&=&k\omega\left(z\frac{d}{dz}+q-\frac{1}{k^2}\right)\pm\Delta\,e^{i\pi\,z\dz}+gz\n
& &+g\, k^k\,z^{-1}\pd_{j=1}^k\left( z\frac{d}{dz}+q-\frac{(j-1)k+1}{k^2}\right).
 \label{k-photon-differentialH}
\eeqa
The corresponding time-independent Schr\"odinger equations are 
\beqa
&&\left\{gk^k\,z^{-1}\pd_{j=1}^k\left( z\frac{d}{dz}+q-\frac{(j-1)k+1}{k^2}\right)+gz\right.\n
&&~~~~~~~\left.+k\omega\left(z\frac{d}{dz}+q-\frac{1}{k^2}\right)
   \pm \Delta\,e^{i\pi\,z\dz}-E_\pm\right\}\psi^\pm(z)=0.  \label{k-photon-Schroedinger}
\eeqa
Here we have written $E_\pm$ since in general the spectra of $H^{(q)}_\pm$ are not the same.
Solutions to the differential equations must be analytic in the whole complex plane and normalizable 
with respect to the norm given in (\ref{norm-for-k-photon}) 
if $E_\pm$ belong to the spectra of $H^{(q)}_\pm$. In other words, we seek solutions of the form
\beq
\psi^\pm(z)=\sum_{n=0}^\infty K_n^\pm(E_\pm)\, z^n, \label{k-photon-series-solution}
\eeq
which converge in the entire complex plane and are elements of ${\cal B}_q$. 

Substituting (\ref{k-photon-series-solution}) into (\ref{k-photon-Schroedinger}), 
we obtain the 3-step recurrence relation,
\beqa
&&K_1^\pm+A_0^\pm\,K_0^\pm=0,\n
&&K_{n+1}^\pm+A_n^\pm\,K_n^\pm+B_n^\pm\,K_{n-1}^\pm=0,~~~~n\geq 1,\label{k-photon-3-step}
\eeqa
where
\beqa
A_n^\pm&=&\frac{\pm(-1)^n\,\Delta-E_\pm+k\omega\left(n+q-\frac{1}{k^2}\right)}{g\,k^{k}\,
   \pd_{j=1}^k\left(n+1+q-\frac{(j-1)k+1}{k^2}\right)},\n
B_n^\pm&=&\frac{1}{k^k\,\pd_{j=1}^k\left(n+1+q-\frac{(j-1)k+1}{k^2}\right)}.
\label{coefficient AB}
\eeqa
The coefficients $A_n^\pm, B_n^\pm$ have the behavior
\beq
A_n^\pm\sim a\,n^\alpha,~~~~ B_n^\pm\sim b\,n^\beta
\eeq
when $n\rightarrow\infty$, where
\beq
a=\frac{\omega}{gk^{k-1}},~~~ \alpha=-k+1,~~~ b=\frac{1}{k^k},~~~ \beta=-k.\label{k-photon-asymptotic}
\eeq
Thus the asymptotic structure of solutions to the $n\geq 1$ part of (\ref{k-photon-3-step}) 
depends on the Newton-Puiseux
diagram formed with the points $P_0(0,0), P_1(1,-k+1), P_2(2,-k)$ \cite{Gautschi67}.

For $k=1$, (\ref{k-photon-3-step})-(\ref{k-photon-asymptotic}) reduce to the corresponding relations obtained in
\cite{Moroz13} for the Rabi model. It was shown that in this case entire wavefunctions 
in an appropriate Bargmann-Hilbert space exist for all model parameters.

\subsection{Wavefunctions and constraints for $\omega$ and $g$ in the $k=2$ case}
For $k=2$ the characteristic equation of the 3-term recurrence relation
is given by $t^2+\frac{\omega}{2g}t+\frac{1}{4}=0$, which  has two solutions
$t_{1,2}=-\frac{\omega}{4g}\pm \frac{1}{2}\sqrt{\frac{\omega^2}{4g^2}-1}$.
Similar to the two-model Rabi case, we now have two cases to consider. 

\vskip.2in
\noindent\underline{\large\bf Case (i)}: $\left|\frac{2g}{\omega}\right|<1$. In this case, we have 
two distinct real roots 
$t_1=\frac{\omega}{4g}\left[-1+\sqrt{1-\left({2g}/{\omega}\right)^2}\right]$,
 $t_2=-\frac{\omega}{4g}\left[1+\sqrt{1-\left({2g}/{\omega}\right)^2}\right]$, and
$|t_1|<|t_2|$. The Perron-Kreuser theorem (i.e. Theorem 2.3 of \cite{Gautschi67})
gives the asymptotic behaviour of two linearly independent
solutions $K^\pm_{n,r}$, 
\beq
\lim_{n\rightarrow\infty}\frac{K^\pm_{n+1,r}}{K^\pm_{n,r}}\sim t_r\,n^{-1},~~~~r=1,2.
\eeq
So $K^\pm_{n,1}$ is the minimal solution and $K^\pm_{n,2}$ is dominant. 
{}From (\ref{k-photon-entireness-test}), for an entire solution in  ${\cal B}_q$ the sum
\beq
\sum_{n=0}^\infty\,|K^\pm_n|^2\,[2(n+q-1/4)]!
\eeq
must converge. Using the ratio test,
\beq
\lim_{n\rightarrow \infty}\frac{|K^\pm_{n+1}|^2\,[2(n+1+q-1/4)]!}{|K^\pm_n|^2\,[2(n+q-1/4)]!}
  =4\,|t_r|^2
\eeq
It is easily seen that $4|t_2|^2>1$. We can show that $4|t_1|^2<1$. Indeed, if we had assumed 
$4|t_1|^2\geq 1$, then we would end up with $\sqrt{1-\left|{2g}/{\omega}\right|}
\geq \sqrt{1+\left|{2g}/{\omega}\right|}$ which is impossible for the non-trivial case $g\neq 0$. 

It follows that the sum $\sum_{n=0}^\infty\,|K^\pm_n|^2\,[2(n+q-1/4)]!$ converges for the 
minimal solution $K^\pm_{n,1}$. Thus the corresponding wavefunctions given by
(\ref{k-photon-series-solution}) are entire functions in ${\cal B}_q$.

\vskip.1in
\noindent\underline{\large\bf  Case (ii)}: $\left|\frac{2g}{\omega}\right|\geq 1$. 
In this case, the two roots $t_{1,2}$ are complex conjugate to each other and $|t_1|=|t_2|=\frac{1}{2}$.
Applying the Perron-Kreuser theorem, we have
\beq
\lim_{n\rightarrow\infty}\,{\rm sup}\,\left(|K^\pm_n|\,n!\right)^{\frac{1}{n}}=\frac{1}{2}
\eeq
for all non-trivial solutions of the 2nd equation of (\ref{k-photon-3-step}). It follows that for a given $\epsilon>0$,
there exists $N(\epsilon)\in {\bf N}$ and an infinite set of $I$ of indices $\ell>N(\epsilon)$
such that $\left(|K^\pm_\ell|\,\ell!\right)^{\frac{1}{\ell}}> \frac{1}{2}-\epsilon$, i.e.
$|K^\pm_\ell|>\frac{(1/2-\epsilon)^\ell}{\ell!}$. Thus we have
\beqa
\sum_{n=0}^\infty\,|K^\pm_n|^2\,[2(n+q-1/4)]! &\geq& \sum_{\ell\in I}\,|K^\pm_\ell|^2\,[2(\ell+q-1/4)]!\n
&>& \sum_{\ell\in I}\,\left(\frac{1}{2}-\epsilon\right)^{2\ell}\,\frac{[2(\ell+q-1/4)]!}{(\ell!)^2}.
\label{inequality2}
\eeqa
Noting that when $\ell\rightarrow\infty$, $\epsilon\rightarrow 0$, we have
\beq
\lim_{\ell\rightarrow \infty}\left(\frac{1}{2}-\epsilon\right)^{2\ell}\,\frac{[2(\ell+1+q-1/4)]!}{(\ell!)^2}
   \neq 0,
\eeq
which means the series on the right hand side of (\ref{inequality2}) diverges. 
Thus by comparison test, the sum $\sum_{n=0}^\infty\,|K^\pm_n|^2\,[2(n+q-1/4)]!$ diverges  
for all non-trivial solutions of the 3-term recurrence relations and 
the 2-photon Rabi model has no entire wavefunctions belonging to ${\cal B}_q$ 
when $\left|\frac{2g}{\omega}\right|\geq 1$. 

\subsection{Energy spectrum in the $k=2$ case}

As in our previous discussion for the regular energy spectrum of the 2-mode Rabi model, the coefficients $K^\pm_n$ will be minimal solutions $K^{\pm min}_n\equiv K^\pm_{n,1}$ iff they satisfy the continued fraction equations
\beq
\frac{K^{\pm min}_{n+1}}{K^{\pm min}_n}=-\frac{B^\pm_{n+1}}{~A^\pm_{n+1}-}\,
  \frac{B^\pm_{n+2}}{~A^\pm_{n+2}-}\,\frac{B^\pm_{n+3}}{~A^\pm_{n+3}-}\,\cdots,   \label{2-photon-continued-fraction}
\eeq
which in turn will require that $E_\pm$ be the roots of
\beq
0=A^\pm_0-\frac{B^\pm_{1}}{~A^\pm_{1}-}\,\frac{B^\pm_{2}}{~A^\pm_{2}-}\,
\frac{B^\pm_{3}}{~A^\pm_{3}-}\,\cdots.   \label{2-photon e-value eqn}
\eeq
Here $A^\pm_n, B^\pm_n$ are given as functions of $E_\pm$ in 
(\ref{2-photon e-value eqn}) with $k=2$. These are transcendental equations whose solutions determine the regular
energies $E_\pm$ of the 2-photon Rabi model.  Only for the denumerable infinite values of $E_\pm$ which are the roots of (\ref{2-photon e-value eqn}), do we get entire wavefunction solutions in ${\cal B}_q$ to the $k=2$ version of the differential equations (\ref{k-photon-Schroedinger}).

\subsection{$k$-photon Rabi model for the $k\geq 3$ case}

We now focus on the $k\geq 3$ case. Let $\sigma$ be the slope of $\overline{P_0P_1}$ and $\tau$ the slope of $\overline{P_1P_2}$
so that $\sigma=\alpha$ and $\tau=\beta-\alpha$. Then $\sigma=-k+1, \tau=-1$, and thus point $P_1$ lies 
below the line segment $\overline{P_0P_1}$ in the Newton-Puiseux diagram.
Applying the Perron-Kreuser theorem (i.e. Theorem 2.3 of \cite{Gautschi67}), we have
\beq
\lim_{n\rightarrow\infty}\;{\rm sup}\,\left(|K^\pm_n|\,(n!)^{\frac{k}{2}}\right)^{\frac{1}{n}}
   =\frac{1}{\sqrt{k^k}}
\eeq
for all non-trivial solutions of the 2nd equation of (\ref{k-photon-3-step}). To have entire solutions which are elements 
of ${\cal B}_q$, the sum on the right hand side of (\ref{k-photon-entireness-test}) must converge. To check if this is the case, we first note that similar to the $k=2$ case, for a given $\epsilon>0$,
there exists $N(\epsilon)\in {\bf N}$ and an infinite set of $I$ of indices $\ell>N(\epsilon)$
such that $\left(|K^\pm_\ell|\,(\ell!)^{\frac{k}{2}}\right)^{\frac{1}{\ell}}> \frac{1}{\sqrt{k^k}}-\epsilon$, i.e.
$|K^\pm_\ell|>\frac{(1/\sqrt{k^k}-\epsilon)^\ell}{(\ell!)^{k/2}}$. So we have
\beqa
\sum_{n=0}^\infty\,|K^\pm_n|^2\,[k(n+q-1/k^2)]! &\geq& \sum_{\ell\in I}\,|K^\pm_\ell|^2\,[k(\ell+q-1/k^2)]!\n
&>& \sum_{\ell\in I}\,\left(\frac{1}{\sqrt{k^k}}-\epsilon\right)^{2\ell}\,\frac{[k(\ell+q-1/k^2)]!}
{(\ell!)^k}.\label{inequality3}
\eeqa
Now
\beq
\lim_{\ell\rightarrow \infty}\left(\frac{1}{\sqrt{k^k}}-\epsilon\right)^{2\ell}\,\frac{[k(\ell+1+q-1/k^2)]!}
  {(\ell!)^k}\neq 0.
\eeq
This means the series on the right hand side of (\ref{inequality3}) diverges.
It follows that the sum on the left hand side of (\ref{inequality3}) diverges for all non-trivial solutions of (\ref{k-photon-3-step}) and the Schr\"odinger equations 
(\ref{k-photon-Schroedinger}) with $k\geq 3$ have no solutions $\psi^\pm(z)$ belonging to ${\cal B}_q$. 

We may thus conclude that for $k\geq 3$ the $k$-photon Rabi model does not have eigenfunctions which are
elements of  ${\cal B}_q$ because they are not normalizable with respect to the norm given in
(\ref{norm-for-k-photon}). In other words, for $k\geq 3$ the Hamiltonian (\ref{k-photonH1}) can not be completely 
diagonalized in the Hilbert space ${\cal H}_b\otimes {\cal C}^2$ due to the non-normalizability of its eigenstates. 
This is in sharp contrast to the $k$-photon Jaynes-Cummings model
which can be completely diagonalized for all $k$ \cite{Lee11}. The impossibility conclusion for the $k\geq 3$ case agrees
with that reached in \cite{Lo98,Ng99,Gorska14} by using different approaches.

\sect{Conclusions}\label{summary}

We have examined the 2-mode and $k$-photon Rabi models based on the application of algebraizations and Bargmann-Hilbert spaces. 
We have seen that the algebraization and parity invariance allow us to decompose 
the total Hilbert spaces into direct sums of independent subspaces, 
thus partially diagonalizing the Hamiltonians of the models by bringing them into block-diagonal forms. 
The block-diagonal sectors can be realized as differential operators 
in the Bargmann-Hilbert spaces. We have investigated the eigenvalues and eigenfunctions of the block-diagonal sectors 
by applying the theory of Bargmann-Hilbert spaces.
We have derived constraints for the frequency $\omega$ and coupling $g$ for the 2-mode and 2-photon Rabi models to be
defined in the Hilbert spaces of entire analytic functions. We have determined the corresponding transcendental equations 
whose roots give the regular energy spectra of the models. 
Furthermore we have shown that the $k$-photon Rabi model with $k\geq 3$ does not have normalizable eigenfunctions with respect to the Bargmann-Hilbert space norm and thus can not be completely diagonalized. 


As shown in the Appendix, the wavefunction expansion coefficients for the 2-mode and 2-photon Rabi models are related to orthogonal polynomials. Thus it is expected that the regular energies of the two models can be determined as the polynomial zeros by a procedure similar to that in \cite{Moroz13,Moroz14}. Work in this direction is in progress and results will be reported elsewhere.

\section*{Acknowledgments}
We would like to thank Alexander Moroz for his interest at the early stage of the work.
Support from the Australian Research Council through Discovery Project DP140101492 is gratefully acknowledged.

\appendix
\section{Orthogonal polynomials}
In this appendix, we show that the expansion coefficients in $\phi^\pm(z)$ (\ref{2-mode-series-solution}) and
$\psi^\pm(z)$ (\ref{k-photon-series-solution}) are related to orthogonal polynomials of infinite degree.

Let us recall the well-known theorem \cite{Chihara78} on relationship between 3-term recurrence
relations and orthogonal polynomials. It states:  the necessary and sufficient condition
for a family of polynomials $\{P_n(x)\}$ (with degree $P_n=n$) in parameter $x$ to form 
an orthogonal polynomial systems is that these polynomials satisfy the 3-term recurrence relation
\beq
P_n(x)=(\beta_n x-\alpha_n)P_{n-1}(x)-\lambda_n P_{n-2}(x),~~~~n\geq 1,
\eeq
where the coefficients $\beta_n, \alpha_n$ 
and $\lambda_n$ are independent of $x$, $\beta_n\neq 0$ and $\lambda_n\neq 0$ for $n\geq 1$.
Then $\{P_n(x)\}$ forms an orthogonal set of polynomials with respect to some weight function.

For the two-mode Rabi case,
the expansion coefficients $S^{\pm}_n(E_\pm)$ defined by the 3-term recurrence relations (\ref{2-mode-3-step}) 
are related to orthogonal polynomials of infinite degree in energy parameters $E_\pm$. To see this, define $P_n^\pm(E_\pm)$ by
\beq
S^{\pm}_n(E_\pm)=\frac{P_n^\pm(E_\pm)}{n!\,(n+2\kappa-1)!}.\label{SP-relation}
\eeq
Then in terms of $P_n^\pm$, (\ref{2-mode-3-step}) becomes
\beq
P_{n+1}^\pm=\frac{1}{g}\left[E_\pm\mp(-1)^n\Delta-2\omega\left(n+\kappa-\frac{1}{2}\right)\right]
  P_n^\pm-n(n+2\kappa-1)P_{n-1}^\pm.
\eeq
It then follows from the above mentioned theorem
that $P_n^\pm(E_\pm)$ are orthogonal polynomials in $E_\pm$ with degree $n$. Thus 
$\phi^\pm(z)$ (\ref{2-mode-series-solution}) are the generating functions for  $P_n^\pm(E)$,
\beq
\phi^{\pm}(z)=\sum_{n=0}^\infty \frac{P_n^\pm(E_\pm)}{n!\,(n+2\kappa-1)!}\,z^n.
       \label{2-mode-series-solution-poly}
\eeq
However, $\phi^{\pm}(z)$ are entire only if $P_n^\pm(E)$ defined in
(\ref{SP-relation}) correspond to the minimal solutions $S^{\pm}_{n, 1}(E)$.

Similarly for the $k$-photon Rabi case, 
the expansion coefficients $K^\pm_n(E_\pm)$ defined by the 3-step recurrence
relation (\ref{k-photon-3-step}) are related to orthogonal polynomials of infinite degree in energy parameters $E_\pm$. This is seen as follows. Define $P_n^\pm(E_\pm)$ by
\beq
K_n^\pm(E_\pm)=\frac{P_n^\pm(E_\pm)}{\prod_{j=1}^k\left[k^n\,\Gamma\left(n+1+q-\frac{(j-1)k+1}{k^2}\right)\right]},
\label{KP-relation}
\eeq
where $\Gamma(x)$ is the gamma function in $x$. In terms of $P_n^\pm(E_\pm)$, (\ref{k-photon-3-step}) become
\beqa
P_{n+1}^\pm&=&\frac{1}{g}\left[ E_\pm\mp (-1)^n\Delta-k\omega\left(n+q-\frac{1}{k^2}\right)\right]P_n^\pm \n
& &-\prod_{j=1}^k\left[k\left(n+q-\frac{(j-1)k+1}{k^2}\right)\right]\,P_{n-1}^\pm.
\eeqa
Thus $P_n^\pm(E_\pm)$ defined by (\ref{KP-relation}) form a set of orthogonal polynomials and the solutions
$\psi^\pm(z)$ (\ref{k-photon-series-solution}) are generating functions for $P_n^\pm(E_\pm)$,
\beq
\psi^\pm(z)=\sum_{n=0}^\infty  \frac{P_n^\pm(E)}{\prod_{j=1}^k\left[k^n\,
   \Gamma\left(n+1+q-\frac{(j-1)k+1}{k^2}\right)\right]} z^n.
\eeq
However $\psi^\pm(z)$ given above have finite radius of convergence (i.e. they are not entire) for $k\geq 3$. 
So we focus on the $k=2$ case in which entire wavefunctions exist. 
Note that in this case the denominator of (\ref{KP-relation}) equals 
$2^n\,\Gamma(n+1+q-1/4)\times 2^n\,\Gamma(n+1+q-3/4) = [2(n+q-1/4)]!$. So the solutions
$\psi^\pm(z)$ for $k=2$ can be expressed as 
\beq
\psi^\pm(z)=\sum_{n=0}^\infty  \frac{P_n^\pm(E_\pm)}{[2(n+q-1/4)]!}\, z^n.\label{2-photon-series-solution-poly}
\eeq
When $P_n^\pm(E_\pm)$ correspond to the minimal solutions $K^\pm_{n,1}(E_\pm)$,
$\psi^\pm(z)$ are entire functions and thus give the wave functions of the 2-photon Rabi model.

\bebb{99}

\bbit{Braak11}
D. Braak, Phys. Rev. Lett. {\bf 107}, 100401 (2011).

\bbit{Moroz12}
A. Moroz, Europhys. Lett. {\bf 100}, 60010 (2012).

\bbit{Chen12} Q.H. Chen, C. Wang, S. He, T. Liu and K.L. Wang, Phys. Rev. A {\bf 86}, 023822 (2012).

\bbit{Zhong13}
H. Zhong, Q. Xie, M.T. Batchelor and C. Lee, J. Phys. A {\bf 46}, 415302 (2013).

\bbit{Moroz13}
A. Moroz, Ann. Phys. {\bf 338}, 319 (2013). 

\bbit{Moroz14}
A. Moroz,  Ann. Phys. {\bf 340}, 252 (2014).

\bbit{Tomka14}
 M. Tomka, O. El Araby, M. Pletyukhov and V. Gritsev, Phys. Rev. A {\bf 90}, 063839 (2014).

\bbit{Batchelor15a}
M.T. Batchelor and H.-Q. Zhou,  Phys. Rev. A {\bf 91}, 053808 (2015) 

\bbit{Batchelor15b}
Z.-M. Li and  M.T. Batchelor,  arXiv:1506.04038v1 [quant-ph].

\bbit{Albert12}
V.V. Albert, Phys. Rev. Lett. {\bf 108}, 180401 (2012).

\bbit{Zhang13a}
Y.-Z. Zhang, J. Math. Phys. {\bf 54}, 102104  (2013).

\bbit{Zhang13c}
Y.-Z. Zhang, arXiv:1304.7827v2 [quant-ph].

\bbit{Braak13}
D. Braak, J. Phys. B {\bf 46}, 224007 (2013).

\bbit{Zhang14}
Y.-Z. Zhang, Ann. Phys. {\bf 347}, 122 (2014).
  
\bbit{Chen14}
Q.-H. Chen, arXiv:1412.8560v1 [quant-ph].

\bbit{Peng15}
J. Peng, Z. Ren, H. Yang, G. Guo, X. Zhang, G. Ju, X. Guo, C. Deng and G. Hao,
J. Phys. A {\bf 48}, 285301 (2015).

\bbit{Lo98}
C. F. Lo, K. L. Liu and K. M. Ng, Europhys. Lett. {\bf 42}, 1 (1998).

\bbit{Ng99}
K. M. Ng, C. F. Lo, and K. L. Liu, Eur. Phys. J. D {\bf 6}, 119 (1999).


\bbit{Gorska14}
K. Gorska, A. Horzela and F.H. Szafraniec, Proc. R. Soc. A {\bf 470}, 20140205 (2014).

\bbit{Bargmann61}
V. Bargmann, Commun. Pure Appl. Math. {\bf 14}, 187 (1961).

\bbit{Barut71}
A.O. Barut and L. Girardello, Commun. Math. Phys. {\bf 21}, 41 (1971).

\bbit{Schweber67}
S. Schweber, Ann. Phys. {\bf 41}, 205 (1967).

\bbit{Gautschi67}
W. Gautschi, SIAM Rev. {\bf 9}, 24 (1967).

\bbit{Leaver86}
E.W. Leaver, J. Math. Phys. {\bf 27}, 1238 (1986).


\bbit{Lee10}
Y.-H. Lee, W.-L. Yang and Y.-Z. Zhang, J. Phys. A {\bf 43}, 185204 (2010).

\bbit{Zhang13b}
Y.-Z. Zhang, J. Phys. A {\bf 46}, 455302 (2013).

\bbit{Lee11}
Y.-H. Lee, J.R. Links and Y.-Z. Zhang, Nonlinearity {\bf 24}, 1975 (2011).



\bbit{Chihara78}
T.S. Chihara, An introduction to orthogonal polynomials, Gordon and Breach, New York, 1978.

\eebb

\end{document}